# Achieving Both High Power and Energy Density in Electrochemical Supercapacitors with Nanoporous Graphene Materials

Hao Yang[1], Santhakumar Kannappan[2], Amaresh S. Pandian[3], Jae-Hyung Jang[2], Yun Sung Lee[3], Wu Lu[1,2]*

[1]Department of Electrical and Computer Engineering, The Ohio State University, Columbus, Ohio 43210, United States.

[2]Department of Nanobio Materials and Electronics, Gwangju Institute of Science and Technology, Gwangju 500-712, Republic of Korea.

[3]Faculty of Applied Chemical Engineering, Chonnam National University, Gwangju 500-757, Republic of Korea.

*Correspondence to: lu.173@osu.edu.

**Supercapacitors, based on the fast ion transportation, are specialized to provide high power, long stability, and efficient energy storage with highly porous electrode materials. However, their low energy density and specific capacitance prevent them from many applications that require long duration. Using a scalable nanoporous graphene synthesis method involving a simple annealing process in hydrogen, here we show graphene supercapacitors capable of achieving a high energy density comparable to what Li-ion batteries can offer, but a much higher power density. Ultra-high specific gravimetric and volumetric capacitances are achieved with highly porous graphene electrodes. Moreover, the supercapacitors assembled with graphene electrodes show excellent stability. Our results demonstrate that by synthesizing graphene materials with an ideal pore size, uniformity, and good ion accessibility, the performance of supercapacitors can be revolutionized.**

Supercapacitors have drawn great attention because of their high charge-discharge rate, long cycle life, outstanding power density, and no short circuit issue that are of concerns with current battery or fuel cell techniques[1-3]. Also known as ultracapacitors or electrochemical capacitors, supercapacitors store energy with electric double-layer (EDL) capacitance (ion adsorption) or pseudocapacitance (surface redox reaction). Although pseudocapacitors using conducting polymers or metal oxides as electrode materials can have a high faradaic capacitive performance, they can't maintain it after long cycling. On the other hand, EDL capacitors with porous carbon electrodes can be charged and discharged as many as one million cycles without performance degradation[4]. Furthermore, the ion transportation in EDL capacitors is faster than the redox reaction in pseudocapacitors, leading to a high charge-discharge rate and power density in EDL capacitors. Currently the major drawback of EDL capacitors is the low energy density generally in the range of 3-5 Wh/kg[5,6], which is one to two orders of magnitude below commercialized lithium-ion batteries (100-275 Wh/kg) [6, 7]. Supercapacitors with increased energy storage but little sacrifice of power density are critically needed for practical applications, such as plug-in hybrid electric vehicles[8], wind turbine energy storage[9], regenerative braking[10] and so on.

To improve the energy density, a porous electrode material with highly developed surface area and high electric conductivity is desired. Carbon materials, due to their low cost, various microtexture and processability, are more attractive than other materials. Activated carbons, carbon aerogels, and carbon nanotubes have been widely used[6]. Activated carbons are the most common materials in EDL supercapacitors nowadays. The reported specific surface area of activated carbon ranges from 500 to 2000 $m^2/g$[11, 12], but only a fraction of it can contribute to energy storage. This is because the activated carbon materials have poor uniformity in pore sizes ranging from micropores (~ 0.3 nm) to macropores[13]. In such materials, the micropores are inaccessible to electrolyte solutions[14] and macropores result in a low surface volume ratio, both of which lead to a low specific capacitance[15]. Carbon nanotubes have a moderate surface area as well as a good conductivity. Supercapacitors based on carbon nanotubes can't exhibit good capacitive performance without adding a pseudocapacitive component[16], and the manufacturing difficulties plus cost further limit their possibility of implementation in actual energy storage devices.

A more promising candidate for the EDL supercapacitor electrode material to overcome the drawback is graphene. Graphene, due to its unique lattice structure, has appealing electrical properties, chemical stability, and high surface area. Ideally a monolayer of $sp^2$ bonded carbon atoms can reach a specific capacitance up to ~ 550 F/g as well as a high surface area of 2675 $m^2$/g, which basically set the upper limits for all carbon materials. With improvements in the exfoliation and reduction techniques of graphite oxide (GO), graphene-based materials can be produced in large-scale at a very low cost. A variety of oxidation-exfoliation methods have been employed to minimize the agglomeration between the graphene sheets, such as chemical reduction[14, 17], microwave irradiation[18], thermal annealing[19, 20], and powerful sonication[21]. Generally chemical reduction methods involve harmful reducing agents and are not suitable for mass production[17]; Thermal annealing can produce high quality graphene sheets but at a cost of very high temperature (~ 1050 °C)[19]; the irradiation and sonication methods serve as pure exfoliation techniques and always require the help of chemical reduction[18, 21]. The reported graphene-based supercapacitors have achieved 85.6 Wh/kg (154 F/g) under hydrazine reduction[14], 70 Wh/kg (166 F/g) with microwave exfoliation[22], and 1.36 mWh/cm³ (276 F/g) by laser scribing[23]. Though tremendous progress has been made, these state-of-the-art values of supercapacitors in energy density is still significantly lower than typical Li-ion batteries[6]. The key challenge to increase the energy storage density of graphene supercapacitors is to develop more effective synthesis approaches offering an ideal pore size, good uniformity, and high ion accessibility.

Differing with previous approaches requiring separate steps for exfoliation and reduction of GO, herein, we show a one-step synthesis route of few-layered graphene sheets in a mesoporous structure with a uniform pore size that can be used as electrode materials for supercapacitors with both high power density and energy density. The synthesis process is based on the exfoliation and reduction reaction of GO with the assistance of hydrogen gas at a relatively low temperature (Fig. 1a, Methods, Supplementary 1). The GO is heavily oxygenated with a basal plane mainly occupied by epoxide (O-C-O), carbonyl (C=O), hydroxyl (C-OH) and carboxyl (COOH) groups[24, 25]. With the presence of heat, oxygen containing groups are reduced by hydrogen to form water vapor between GO layers. When water vapor accumulates, a high pressure is generated to break the agglomeration, where the reduced GO is converted into hydrogen-annealed graphene (HAG). The process can be visually observed with the color change from brown to black (related to reduction) and large volume expansion (related to exfoliation) (Supplementary Fig. 1). The reduction and exfoliation take place simultaneously in this approach avoiding using toxic chemicals and high temperature thermal processing. The field emission scanning electron microscopy images (Fig. 1b, and Supplementary Fig. 3a, b) show the typical cross section of HAG surface. They reveal the successful exfoliation under which a large area of restacking layers has been peeled off so that it shows obvious gaps between them. The transmission electron microscopy images (Fig 1 c, b, and Supplementary Fig. 3c, d) show the morphology of HAG powder, where the agglomeration is very limited. Hence, supercapacitors can be made with lightweight and highly porous HAG electrodes.

To characterize the surface area and porosity, the study of nitrogen isothermal adsorption experiment was performed at liquid nitrogen temperature. It shows a type IV adsorption isothermal curve with a hysteresis loop generally exhibited by mesoporous solids[26] (Fig. 2a). The hysteresis loop between adsorption and desorption along with a sharp fall at a relatively high pressure implies the slit geometry of major mesopores[27], which appears to be in good agreement with the electron microscopy images (Fig. 1b, c, d and Supplementary Fig. 3). The HAG demonstrates a Brunauer-Emmett-Teller (BET) specific surface area of 410 m$^2$/g, which is extracted from the linear region of $1/[V_a(p_0/p - 1)]$ versus $p/p_0$ in the classical BET range of 0.05-0.3[28] (Inset of Fig. 2a). The pore size distribution (Fig. 2b) is extracted by Barret-Joyner-Halenda (BJH) method. A peak corresponding to a pore volume of 2.46 cm$^3$/g occurs at a pore size of 4.27 nm, and the high peak sharpness suggests the good uniformity of pore size distribution.

The X-ray diffraction features of GO show an intense (002) peak occurring at 12.06° with a d-spacing of 7.31 Å. Upon the hydrogen treatment, GO diffraction peak disappears and a new peak with reduced intensity and broadened width appears around 24.44° for HAG (Fig. 3a). The interlayer spacing of the reduced product decreases to a much smaller value of 3.64 Å which suggests the removal of oxygen containing groups by reacting with hydrogen molecules under heat. Compared with the interlayer distance of graphite ~ 3.35 Å[29], the shift implies that there are some residual functional groups left. The appearance of broad peak is an indication of a loss of order in HAG.

After Lorentzian fitting of Raman spectrum of HAG (Fig. 3b), the D-band caused by disordered structure can be found approximately at 1350.96 cm$^{-1}$. It's pretty common that the oxidation approaches generate a certain amount of defects in graphene sheets. The G-band corresponding to E$_{2g}$ mode at the center of Brillouin zone locates at ~ 1582.85 cm$^{-1}$ [30]. The intensity ratio of D/G bands is often used as prediction of defects in samples. The ratio of HAG D-band and G-band, $I_D/I_G$, is around 0.86 less than 1.0, indicating the restoring of the π-conjugated structure[30].

In the X-ray photoelectron spectrum (Fig. 3c), the C/O ratio increases from ~ 2 to ~ 5 with O1s peak reduced significantly from GO to HAG. By performing Gaussian-Lorentzian fitting, the intense C-C peak is observed at a binding energy of 284.6 eV showing the sp$^2$ bonding. The broad tail towards higher binding energy up to 296 eV is because of the contributions of various carbon bonding configurations. Those multiple peaks at 285.99, 287.37, 288.93 eV are typically assigned to oxygen containing groups C-OH (or C-O-C), C=O and COOH respectively[31]. The existence of C-O-C which is widely seen in the graphene oxide system has a similar C1s binding energy to C-OH[32]. Another confirmation of the reduction mechanism is obtained by the Fourier transform infrared spectroscopy. In HAG, the broad and intense peak at 3433 cm$^{-1}$ and 1637 cm$^{-1}$ [33] correspond to the OH stretching in water molecules indicating that there are lots of H$_2$O generated after reduction. This phenomenon is consistent with our hypothesis of hydrogen reduction with water vapor as the reaction product. Moreover, the C=C stretching is observed at 1560 cm$^{-1}$. Other than the two peaks, the peaks at 1719 cm$^{-1}$, 1383 cm$^{-1}$, and 1190 cm$^{-1}$ are related to the C=O, C-O, and -OH stretching in COOH, respectively[34]. The weak intensities of these peaks suggest the efficient removal of those functional groups. The asymmetric and symmetric stretching peaks of –CH$_2$ at 2918 cm$^{-1}$ and 2851 cm$^{-1}$ reveals the restoration of carbon basal planes owing to hydrogen reduction[33, 35].

The prepared HAG electrodes show a highly porous structure with which a high percentage of the surface area is accessible for electrolyte ions (Supplementary Fig. 5). To evaluate the electrochemical performance of HAG electrodes, supercapacitors were assembled in symmetrical cell geometry (Fig. 4a and Supplementary Fig. 2). The devices were measured with ionic liquid 1-Ethyl-3-methylimidazolium tetrafluoroborate (EMIMBF$_4$) and LiPF$_6$ electrolytes, respectively. When capacitance C is fixed with swept voltage, current will retain constant value because of the relation $I(t) = C(dV/dt)$[36]. Therefore, the resulted cyclic voltammetry (CV) curves of ideal capacitors are expected to display rectangular current response, if the same scan rate $dV/dt$ is applied in forward and reverse directions of the sweep. The CV curves of HAG electrodes in EMIMBF$_4$ and LiPF$_6$ are nearly rectangular from scan rate 5 mV/s to 100 mV/s indicating the capacitive behavior (Fig. 4b and Supplementary Fig. 4a). The ionic liquid EMIMBF$_4$ has moderate high conductivity and can work under a high voltage up to 4.3 V[14, 37], which brings significant benefits in improvement of energy density.

Nyquist plots of electrochemical impedance spectroscopy (EIS) (Fig. 4c) show a semicircle in high frequency region and a straight line in the low frequency region. Compared with LiPF$_6$, the more vertical line of ionic liquid proves a nearly ideal capacitive response. The equivalent series resistance (ESR) obtained from the x-intercept of Nyquist plot is 4.26 ohm. The low ESR value is mainly responsible for the high power density of mesoporous graphene supercapacitors despite of the high viscosity of ionic liquid.

The galvanostatic charge-discharge measurements were taken at various current densities (Fig. 4d and Supplementary Fig. 4b). The nearly linear discharging curves illustrate an ideal

EDL capacitor performance which can be found in $dV/dt = I/C$. The voltage drop at the beginning of discharging is due to the voltage loss across the ESR.

The typical specific capacitance of graphene supercapacitors in ionic liquid is in the range of 100-250 F/g at a current density of 1 A/g[14]. The experimental specific capacitance values of the HAG electrode in ionic liquid extracted from discharge curves are 306.03, 256.19, 239.47, 216.99, 180.30, and 146.77 F/g at current densities of 1, 2, 2.5, 4, 5, and 8 A/g, respectively (Fig. 4e). And the specific capacitance values in LiPF$_6$ are 111.10, 95.33, and 90.09 F/g at 1, 2.5, and 4 A/g, respectively (Fig. 4e). Note that these values are based on the total weight of electrode materials. The specific capacitances of LiPF$_6$ at the same current densities are smaller as a result of the natural properties and different breakdown voltage limits of electrolytes. Based on the Gouy-Chapman-Stern model, the Debye lengths for those two electrolytes, and the specific surface area of HAG, we have calculated the theoretical EDL capacitances. In EMIMBF$_4$ and LiPF$_6$, the ideal capacitance values are 346.90 F/g and 136.21 F/g, assuming the full ion accessibility of the material surface, which are just slightly higher than the measured results at low current density, indicating a high ion accessibility of synthesized HAG materials (88% for ionic liquid and 82% for LiPF$_6$, Supplementary 8). The high ion accessibility is a result of the mesoporous structure of HAG. Because the pore size of HAG is few times larger than the size of electrolyte ions, ions can easily accommodate inside the HAG electrodes leading to better electrolyte accessibility and easy ion transportation. This porous surface allows electrolyte to access the interior region even when compressed to electrodes in supercapacitors.

To reveal the energy storage performance, the Ragone plots of supercapacitors in EMIMBF$_4$ and LiPF$_6$ are presented in Fig. 4f and Supplementary Fig. 6, respectively. An ultra-high energy density of 148.75 Wh/kg (50.41 mWh/cm$^3$) at 1 A/g was achieved with ionic liquid at room temperature. For comparison, previous reported gravimetric energy density values of graphene supercapacitors are between 20-90 Wh/kg at 1A/g (Supplementary Table 1)[14, 22, 23, 38]. The volumetric energy density values are in the range of 1 to 30 mWh/cm$^3$ for supercapacitors with no special compression methods applied (Supplementary Table 1)[14, 22, 23]. In addition, the highest power density in ionic liquid is 41 kW/kg obtained at 8 A/g while the highest value in LiPF$_6$ is 47 kW/kg at 2.5 A/g. The stability of HAG electrodes is confirmed by long charge-discharge test. After charging and discharging for 10,000 cycles (Fig. 5), 88% of the initial specific capacitance was retained in ionic liquid. For demonstration, one of the HAG supercapacitors was packaged in a coin cell and charged to 4V. It was then connected with discharging circuits to show its capability of releasing the stored energy to power up various light emitting diodes (Supplementary Video 1).

The outstanding electrochemical performance of the HAG supercapacitors meets the power, energy, and durability requirements of many applications that batteries and fuel cells can't[7]. In addition, the material synthesis presented in this paper is a low-temperature, environmentally friendly, time-efficient, and low cost process which is readily scaled up for industrial manufacturing. During the synthesis, it only involves a one-step treatment to exfoliate and reduce the GO material and comes with no following activation process to improve the porosity and ion accessibility. The appealing performance of this graphene supercapacitor technology could bring in new electrochemical energy storage applications with both high power and energy requirements.

**METHODS**

GO was prepared by oxidizing the graphite powder with modified Hummer's method. After vacuum-drying, GO powder was put into a quartz furnace under flowing hydrogen gas at a pressure of 110 torr for 3 minutes. The morphology of the synthesized material was characterized by field emission scanning electron microscopy (FESEM, HITACHI S-4700, Japan) and high-resolution transmission electron microscopy (HRTEM, TECNAI F20), respectively. The measurement of nitrogen adsorption isothermal was done at 77 K by low temperature nitrogen adsorption surface area analyzer (ASAP 2020, Micromeritics Ins, USA). The diffraction patterns were tested by a high resolution X-ray diffractometer (HR-XRD, Rigaku, Japan) with Cu Ka radiation (k = 1.54056 Å). Raman spectra were obtained with Horiba Jobin-Yvon, France with a 514 nm $Ar^+$ ion laser excitation source at 10 mW. X-ray photoelectron spectroscopy was measured on a MULTILAB 2000 system (SSK, USA). The Fourier transform infrared (FTIR) study was carried out to examine the vibrational characteristics of graphene using an IR Prestige-21 system (Shimadzu, Japan). CV was carried out with a CHI760D electrochemical analyzer (CH Instruments, Inc., USA) at a scan rate from 5 mV/s to 100 mV/s. EIS was carried out within a frequency range of 100 kHz to 0.1 Hz at an open-circuit potential with an a.c. amplitude of 10 mV with a Zahner Electrochemical Unit 1M6e EIS system (Zahner, Germany). Galvanostatic charge–discharge cycling of the cells was performed at a current density from 1 A/g to 8 A/g with a battery tester (NAGANO, BTS-2004H, Japan).

**Acknowledgments** This work is partially supported by the World Class University program by Korean National Research Foundation (Project No. R31-10026) and National Science Foundation.

**Author Contributions** W.L., Y.S.L., and J.H.J. conceived the experiments. H.Y. and S.K. synthesized the material. H.Y. performed the analysis and calculation based on the discussion with W.L. and S.K.. A.S.P. prepared the electrode and assembled the supercapacitor cell. H.Y. wrote the manuscript. W.L. and H.Y. revised and finalized the manuscript.

**Figure Captions:**

Fig. 1 (a) Schematic demonstrating the process of one step exfoliation and reduction process of GO with low temperature Hydrogen annealing. During this process, hydrogen gas reacted with oxygen related groups forming water vapor and exfoliate the restacking layers. (b) A FE-SEM image of the HAG piece. (c) A TEM image of a HAG sheet. It reveals the typical morphology of those graphene materials. (d) A higher resolution TEM image of the edge of a HAG sample.

Fig. 2 (a) Nitrogen adsorption/desorption plot at 77.4 K of the HAG sample. Inset gives the plot of $1/[V_a(p_0/p-1)]$ versus $p/p_0$ in the classical BET range of 0.05-0.3, based on which the specific surface area is extracted. The type IV adsorption isothermal curve with a hysteresis loop suggests the near cylindrical or slit geometry of major pores. (b) Pore-size distribution of the HAG sample.

Fig. 3 (a) XRD spectra of GO and HAG samples showing the interlayer spacing decrease after the exfoliation and reduction. (b) Raman spectrum of the HAG sample. (c) XPS C1s spectrum of the HAG sample and the inset is the comparison between GO and HAG. (d) FTIR spectrum of the HAG powder.

Fig. 4 (a) Schematic of the symmetrical supercapacitor structure. From the left to middle are cell cap, HAG electrode, electrolyte and separator. (b) CV curves of HAG electrode in EMIMBF4 at different scan rates. (c) Nyquist plot from EIS measurement. (d) Galvanostatic charge-discharge curves of supercapacitor in EMIMBF4 for different current densities. (e) Specific capacitance versus current density for HAG supercapacitors. (f) Ragone plot of the supercapacitor in EMIMBF4. The record high energy density of 148.75 Wh/kg is achieved at 1A/g.

Fig. 5 The HAG supercapacitors show excellent stability for 10,000 cycles.

Fig. 1

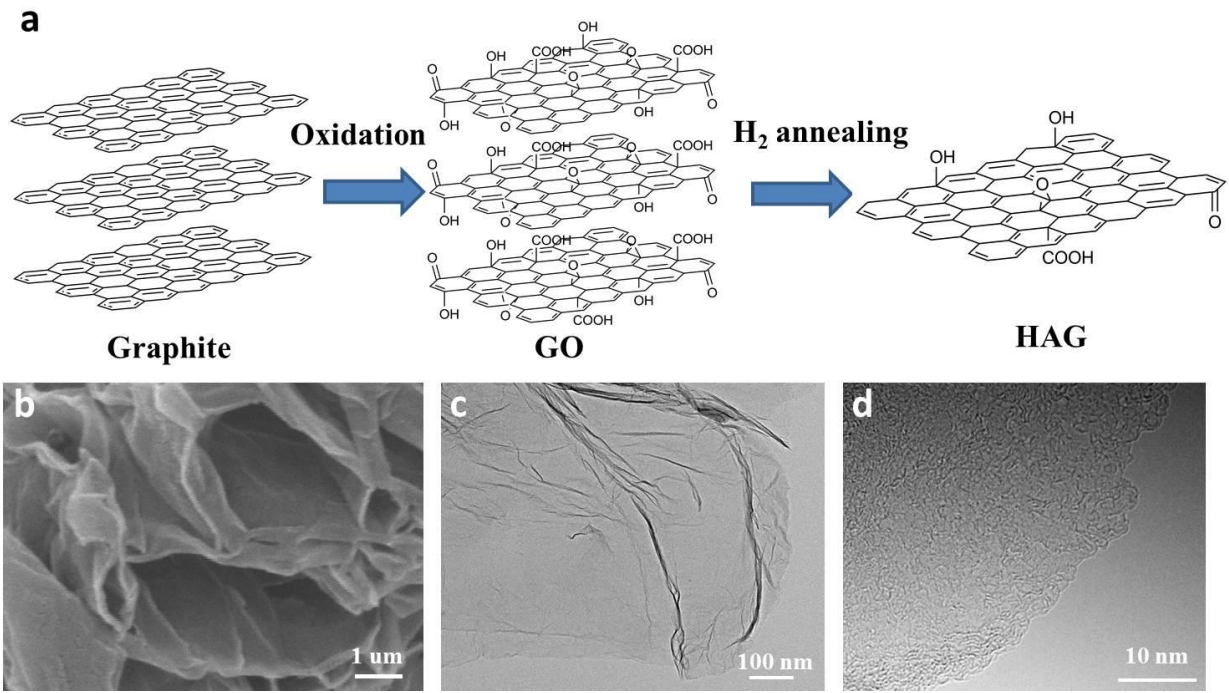

Fig. 2

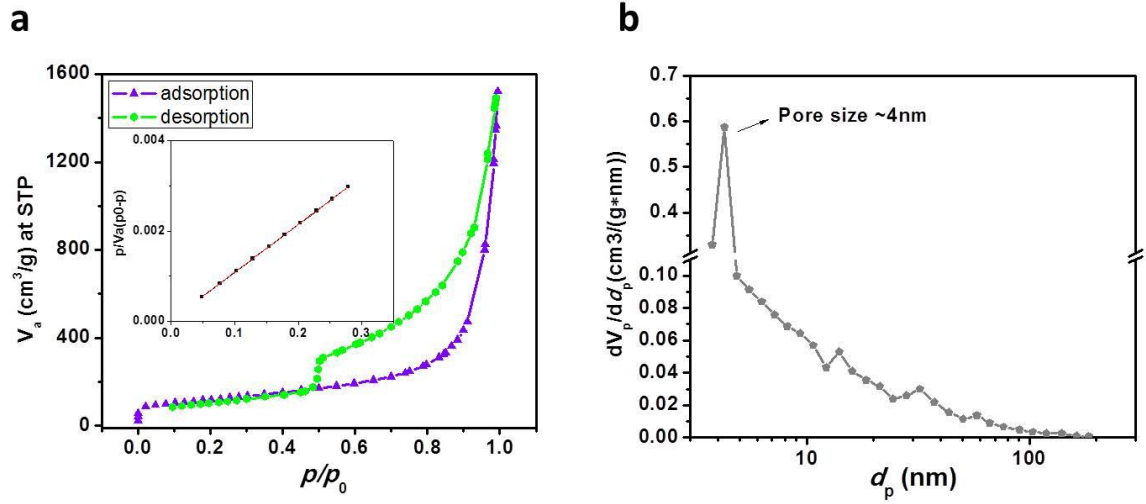

Fig. 3

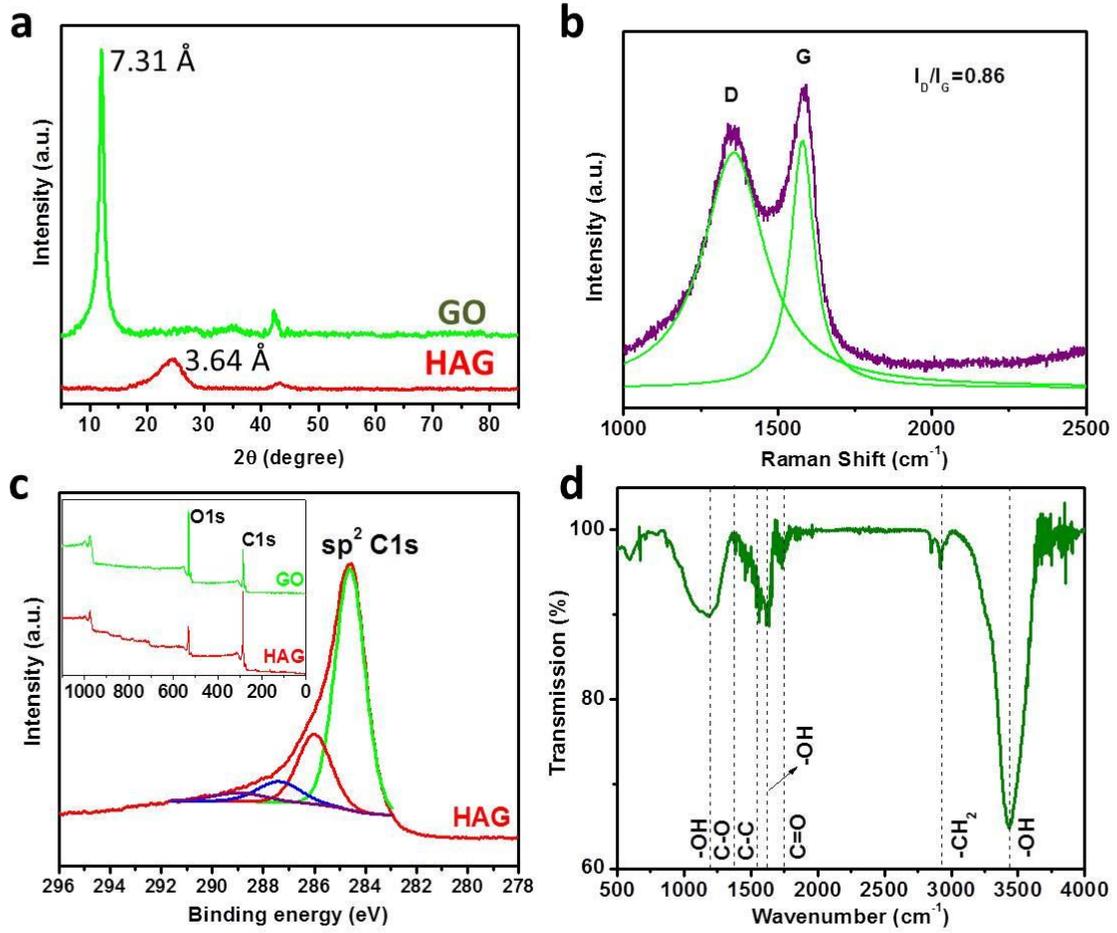

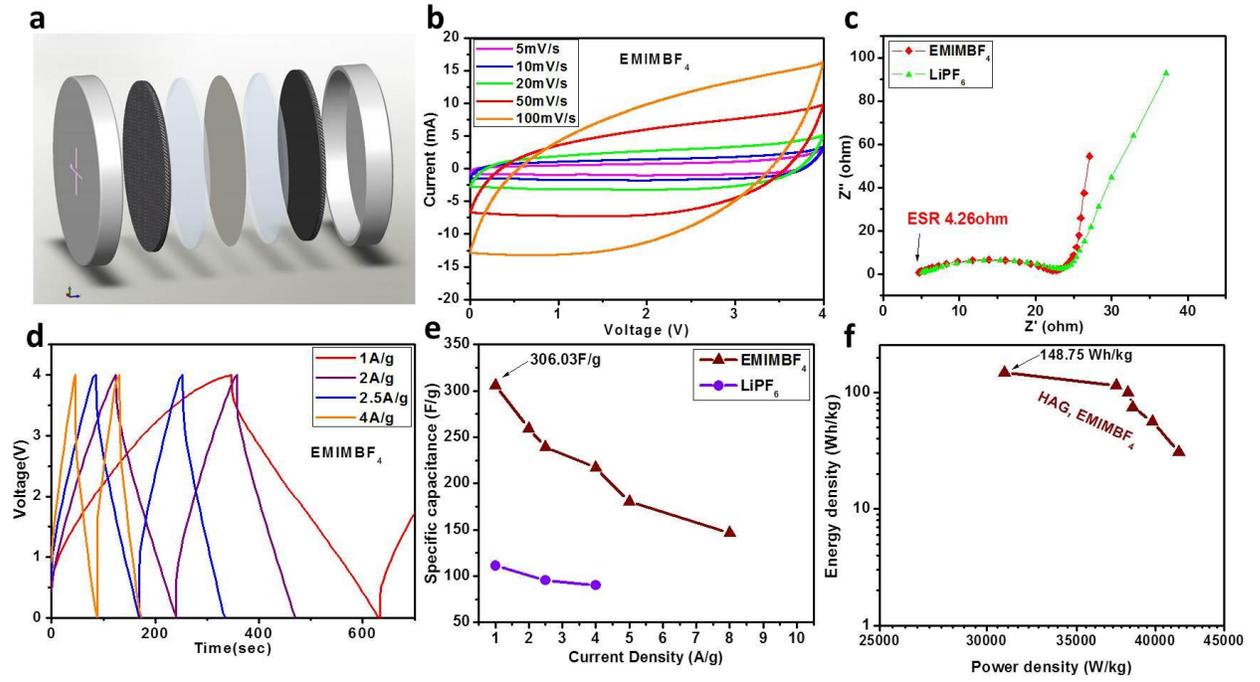

Fig.5

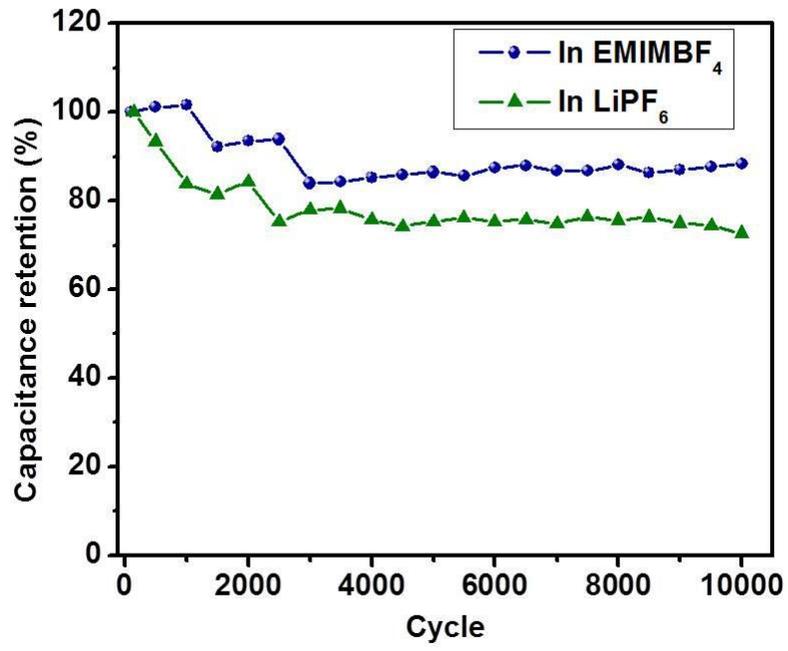